\documentstyle[preprint,eqsecnum,aps]{revtex}
\begin{document}
\tighten
\draft
\preprint{UCSBTH-97-11, NSF-ITP-97-053, gr-qc/970xxx }

\title{Quantum Pasts and the Utility of 
History\thanks{Talk presented at {\sl The 
Nobel Symposium: Modern Studies of Basic Quantum Concepts and Phenomena},
Gimo, Sweden, June 13-17, 1997}}

\author{James B.
Hartle\thanks{e-mail: hartle@cosmic.physics.ucsb.edu}}
\vskip .13 in
\address{Institute of Theoretical Physics\\
and Department of Physics,\\
University of California,\\
Santa Barbara, CA
93106-4030}

\maketitle

\date{\today}
\begin{abstract}
{From} data in the present we can predict the future and retrodict the
past. These predictions and retrodictions are for histories --- most
simply time sequences of events. Quantum mechanics gives probabilities
for individual histories in a decoherent set of alternative histories.
This paper discusses
several issues connected with the distinction between prediction and
retrodiction in quantum cosmology: the difference between classical and
quantum retrodiction, the permanence of the past, why we predict the
future but remember the past, the nature and utility of reconstructing
the past(s), and information theoretic measures of the utility of
history.

\end{abstract}

\setlength{\baselineskip}{.3in}


\tighten
\setcounter{footnote}{0}
\section{Introduction}
\label{sec:I}

Up here on length scales above the Planck length, where
the notion of an approximately classical spacetime makes sense, the
world is four-dimensional with three spacelike dimensions and one for
time.  Classically we most accurately describe the world in terms of the
four-dimensional concepts of points, world lines, and field
configurations in spacetime.  Yet we can divide our
subjective experience up into past, present, and future.  From our point
of view these have very different properties: We know the present,
remember the past, and predict the future.

The processes of predicting the future and retrodicting the past
 are familiar and elementary
in classical physics.   However, fundamentally the world is not classical
but quantum mechanical.  The process of prediction --- the prediction
 of a measurement outcome, for instance  --- is also familiar in
quantum mechanics.  The
quantum past, however is more of a foreign country.  This essay is concerned
with the past in quantum mechanics.  What is the nature of quantum
mechanical retrodiction? What is the origin of the differences between
past and future?  Most importantly, why are we interested in the past at
all?  It's over. What is the utility of history?

\section{Prediction and Retrodiction in Quantum Mechanics}
\label{sec:II}

\subsection{Classical Prediction and Retrodiction}
\label{sec:II.A}

Classically, prediction and retrodiction are symmetrically related if we
presume that the fundamental equations of motion are time reversal
invariant. 
{From} present values for the positions and velocities
of the molecules in a gas we may predict their future positions and
velocities and retrodict their past ones by using the deterministic
equations of motion. 
Effective classical equations for coarse-grained quantities such as the
density of the gas may
exhibit dissipation, chaos, and history dependence which distinguish
prediction from retrodiction, but {\it in principle}, if present data
contained enough information, the processes of prediction and
retrodiction would be symmetrically 
related. That is not the case in quantum mechanics.

\subsection{The Quantum Mechanics of a Closed System}
\label{sec:II.B}

We consider the quantum processes of prediction and retrodiction 
from the most
comprehensive perspective --- the quantum mechanics of a closed system, most
generally the universe as a whole. (See, {\it e.g.}~\cite{Omn94}). 
We neglect quantum gravity to keep
the discussion manageable. This is an excellent approximation for many
useful purposes for any time later than
$10^{-43}$sec after the big bang. In this
approximation the universe may be thought of as a collection of matter
fields inside a large, perhaps expanding, box in fixed background
spacetime geometry. The two fundamental inputs
that specify a closed quantum mechanical system are its Hamiltonian $H$ and
initial quantum state $|\Psi\rangle$.  We now very briefly review how
these lead to predictions and retrodictions.

 General objectives of quantum theory are the probabilities of
individual time histories in sets of alternative coarse-grained histories
of the universe.  Examples are
the probabilities of a set of possible orbits 
of the earth
around the sun. Such histories are said to be coarse grained because
they do not specify the co\"ordinates of every particle in the universe
but only those of the center of mass of the sun and earth and these only
crudely and not at every time.

The simplest sets of alternative histories are alternatives at a moment
of time.  These can always be reduced to a set of yes/no alternatives.
For instance, questions about the position of the earth's center of mass
can be reduced
to questions of the form: ``Is the earth's center of mass
in this region --- yes or
no?'',  
``Is it in that region --- yes or no?'', {\it etc.}
 A set of yes/no
alternatives at a moment of time, say $t=0$, is represented by a set of
orthogonal projection operators $\{P_\alpha\},\, \alpha = 1,2,\cdots$ ---
one projection operator for each alternative.   These projection
operators must satisfy 
\begin{equation}
\sum\nolimits_\alpha P_\alpha = I, \quad {\rm and} \quad P_\alpha P_\beta = 0,
\ \alpha \not= \beta, 
\label{twoone}
\end{equation}
showing that they represent an exhaustive set of exclusive
alternatives.   
In the Heisenberg picture that we shall employ, the same
set of alternatives at a later time is represented by a different set of
operators related to the first by
\begin{equation}
P_\alpha(t) = e^{iHt} P_\alpha e^{-iHt}\ .
\label{twotwo}
\end{equation}
(Here and throughout we use units where $\hbar=1$.)

A set of alternative histories may be specified by giving 
sets of alternatives at a series of times $t_1< t_2 < \cdots < t_n$. We
denote the corresponding sets of projections by $\{P^1_{\alpha_1}(t_1)\},
\cdots, \{P^n_{\alpha_n} (t_n)\}$. An individual history corresponds to
a particular sequence of alternatives $(\alpha_1, \cdots, \alpha_n)\equiv
\alpha$ and is represented by the corresponding chain of projections
\begin{equation}
C_\alpha = P^n_{\alpha_n} (t_n) \cdots P^1_{\alpha_1} (t_1) \, .
\label{twothree}
\end{equation}
Such a set of histories is said to be {\it coarse-grained} when there
is not a set of alternatives at each and every time or when
the projections are not  not one-dimensional
onto complete sets of states. 
For a set of histories describing alternative orbits of the earth,
the $\{P^k_{\alpha_k}(t_k)\}$ might be sets of projections onto an
exhaustive set of ranges of position of the earth's center of mass
at different times $t_k$. A coarse-grained individual orbit is a sequence of
these ranges at a series of times. 

When the $C_\alpha$ act on the initial state they give branch state
vectors $C_\alpha|\Psi\rangle$ for each of the alternative histories in
the set.
The probabilities of the individual histories are
\begin{equation}
p(\alpha) = \left\Vert C_\alpha |\Psi\rangle\right\Vert^2\ .
\label{twofour}
\end{equation}
Eq.(\ref{twofour}) displays an important asymmetry in time. There is
$|\Psi\rangle$ on one end of history
and nothing on the other. That is the arrow of time in quantum theory.
This is the only fundamental time asymmetry and all other 
``arrows of time'' must arise from it and particular properties of the
$|\Psi\rangle$ of our universe. Similarly this is the only source for
the distinction between past and future. The question is not ``Where
does the distinction between past and future come from?'' but rather
``How does the distinction between past and future arise from
special properties of the initial condition of the universe
in conjunction with the quantum mechanical arrow of time?''

However, it is a convention that the state $|\Psi\rangle$ is an {\it initial
} condition at a moment earlier than all others in history. Field
theory is $CPT$ invariant, and by making use of a $CPT$ transformation
the time order of the the operators and $|\Psi\rangle$ could be inverted
so that $|\Psi\rangle$ was a final condition. That convention however
is in conflict with ordinary language and therefore risks confusion. 
Thus in this paper, by ``past'' we mean the times on the side of the
present which is closest to the initial condition. 

Quantum mechanics can be formulated time neutrally  with
initial and final density matrices $\rho^i$ and $\rho^f$ entering
symmetrically into the expressions for probabilities \cite{TNQMsum}.
So formulated,
quantum mechanics contains no intrinsic direction of time, and
all arrows of time arise from asymmetries between the initial and
final conditions. It appears that a final condition of indifference,
$\rho^f=I/Tr(I)$ and a special initial condition, possibly pure,
$\rho^i=|\Psi\rangle\langle\Psi|$, are consistent with known data
\cite{Laf93,Cra95}. In that case the time neutral formulation reduces
to the one used here summarized by (\ref{twofour}). 

Probabilities cannot be consistently assigned to {\it every} set of
alternative histories of a closed system because of quantum mechanical
interference.  In the two-slit experiment, an electron emitted by an
electron gun can pass through either of two slits in a screen on its way
to detection at a further screen.    It would be
inconsistent to assign probabilities to the two histories distinguished
by which slit the electron goes through.  The probability to arrive at a
definite point on the detecting screen would not be the sum of the
probabilities to arrive there by going through each of the two different slits
because of interference.
The sets of histories to which probabilities {\it can} be assigned are
those which have negligible interference between their individual
members, that is
\begin{equation}
\left\langle\Psi|C^\dagger_\alpha C_\beta |\Psi\right\rangle \approx 0\ ,
\ \alpha\not=\beta\ .
\label{twofive}
\end{equation}
Such sets of histories are said to {\it decohere} and (\ref{twofive}) is a
decoherence condition.  The probabilities (\ref{twofour}) satisfy the 
additivity properties required of probabilities as 
a result of (\ref{twofive}).  As a consequence 
(\ref{twofour}) is a consistent assignment of probabilities for
decoherent sets. The sets of alternative histories which can
be assigned probabilities are thus determined through (\ref{twofive}) by
the initial condition of the universe. 

As an example of a mechanism of decoherence, think of a millimeter-size
dust grain in a superposition of two positions deep in intergalactic
space \cite{JZ85}. Consider alternative histories of the position of this particle
at a series of a few times.  Were the grain truly isolated this
situation would be analogous to the two slit experiment and alternative
histories of its position would not decohere.  However, in our universe,
even deep in space, about $10^{11}$ cosmic blackbody photons scatter from
it every second.  Through these interactions, states of
 the seemingly isolated dust
grain become strongly correlated with states of the radiation. 
Two states of position
differing by a millimeter become correlated with nearly orthogonal
states of the radiation after about a nanosecond. The branch state
vectors corresponding to different suitably coarse-grained histories of
position thereby become nearly orthogonal, eq.~(\ref{twofive}) is
satisfied, and decoherence achieved.
Mechanisms such as this are widespread in our universe and cause the
habitual decoherence of the kinds of quasiclassical variable we so often
find it useful to follow.  In what follows we shall always assume that
we are dealing with a decoherent set of histories.

\subsection{Prediction and Retrodiction}
\label{sec:II.C.}

{From} known present data we aim to predict the future and
retrodict the past. The exact nature of this present data is not
important for the subsequent discussion.  It could be the memories of an
individual observer, the records of a collection of them, or an
alternative having nothing to do with observers such as the present
size of the universe. Neither does the data from which predictions and
retrodictions are inferred have to be at one moment of time, and indeed
typically it is
not. However, to simplify the exposition, we shall assume that it is;
the generalization to data at several times should be clear.

 Probabilities of future alternatives conditioned on
present data are relevant for prediction. Probabilities for past
alternatives conditioned on the same present data are required for
retrodiction. 
These are easily constructed from the joint probabilities (\ref{twofour})
of histories that include present data, as we now show.

Let $t_0$ be the time of the present and denote the projection
corresponding to present data by $P_{pd(t_0)}(t_0)$. The awkward
subscript $pd(t)$ is necessary because present data will
change in time as new data is acquired and old data lost. The 
operator $P_{pd(t_0)}(t)$, evolved from $t_0$ according to
(\ref{twofour}), would represent that same alternative as
$P_{pd(t_0)}(t_0)$ at time $t$, but that is not present data at 
time $t$. However, since we only use the operator $P_{pd(t_0)}(t_0)$,
we shall often abbreviate it as $P_{pd(t_0)}$ or even $P_{pd}$. 

The conditional
probability for a chain of alternatives $(\alpha_n,\cdots,\alpha_1)$ at
times $t_n>\cdots>t_1$ all to the future of $t_0$ is
\begin{equation}
p\left(\alpha_nt_n,\cdots,\alpha_1t_1|pd \,t_0\right) = \frac{p\left(\alpha_n
t_n,\cdots,\alpha_1t_1,\, pd\, t_0\right)}{p(pd\, t_0)}\ .
\label{twosix}
\end{equation}
Using (\ref{twofour}), and assuming the set of histories is decoherent,
the joint probabilities in the numerator and denominator may be
evaluated from the relevant chains of projections 
and the state $|\Psi\rangle$.
The results are conditional probabilities for prediction.

Similarly for retrodiction.  The conditional probabilities for a history
of alternatives $(\alpha_1,\cdots\alpha_n)$ at times
$\tau_1>\cdots>\tau_n$ all {\it prior} to $t_0$ is
\begin{equation}
p\left(\alpha_1\tau_1, \cdots, \alpha_n\tau_n|pd\, t_0\right) = 
\frac{p\left(\alpha_1\tau_1,\cdots,\alpha_n\tau_n, pd
\, t_0\right)}{p(pd\, t_0)} \, , 
\label{twoseven}
\end{equation}
where the chain relevant for computing the numerator is
\begin{equation}
P_{pd (t_0)}(t_0) P^1_{\alpha_1}(\tau_1)\cdots P^n_{\alpha_n}(\tau_n)\ .
\label{twoeight}
\end{equation}
Eq.~(\ref{twoseven}) gives conditional probabilities for retrodiction.

For example, to predict or retrodict the orbit of the earth about the sun
$P_{pd(t_0)}$ might be the projection on a present record of
previous observations of the earth's position to certain accuracies,
 and $P^k_{\alpha_k}(t_k), k=1,\cdots,n$ might be projections
on a series of future or past positions to given accuracies. For typical
accuracies to which the position of the earth's center of mass
is determined, we expect there to be a conditional probability
near unity for classical orbits consistent with present data. 
  
\subsection{Retrodiction in the Approximate Quantum Mechanics of
Measured Subsystems.}
\label{sec:II.D.}

The Copenhagen formulation of quantum mechanics is an approximation to
the more general quantum mechanics of closed systems. 
It is appropriate when dealing with measurement situations in which a
subsystem is approximately isolated from the rest of the universe
 and evolves independently
except for occasional interactions with a measuring apparatus\footnote{
See {\it e.g.} \cite{Har91a}, Section II.10.}.  In ideal situations,  
the decoherence of the
alternative histories of the registrations by the apparatus may be
approximated as exact, and the probabilities for the outcomes of
measurements calculated from a formula similar to (\ref{twofour}) but
{\it involving the Hilbert space of the measured subsystem alone.} The
change in focus from the Hilbert space of the $10^{80}$ particles in the
visible universe to say, the Hilbert space of the few particles involved
in a typical experiment is an approximate simplification of
overwhelming practical advantage. 

There is an obstacle to using the Copenhagen probabilities to retrodict
the past of a subsystem. 
The output of the Copenhagen approximation are the
probabilities of the outcomes of measurements.  But, there is not
enough information in the present state of a subsystem to fix what
past measurements were performed on it and when they were performed. 
The influence of these
measurements on the history of the subsystem's state is generally
non-negligible.
To retrodict, the information about the nature and times of measurements must either be
supplied, or perhaps itself retrodicted from a separate
classical physics governing the measuring apparatus.

In this practical sense it is not possible to retrodict the past from
the present data in the Copenhagen formulation of quantum mechanics,
except for those situations where evolution
is negligibly perturbed by
external measurements as can be arranged in classical physics.  A similar 
obstacle exists for prediction.  This appears less
serious because as observers we can control the nature of time of {\it
future} measurements whereas those in the past seem beyond our reach.
True quantum mechanical
retrodiction is only possible for a closed system.

\section{Quantum Pasts}
\label{sec:III.}

The reconstruction of the past is most honestly viewed in the context of
retrodiction from present records.  We retrodict the date 55BC for the
first Roman invasion of Britain on the basis of present textual records.  We
use present observations of the sun and moon to reconstruct
their past trajectories. We use fossil records to estimate that the
probability is high that dinosaurs roamed the earth 250 million years
ago. We infer that matter and radiation were in thermal equilibrium at
the beginning of the universe from the present values of the Hubble
constant, mean mass density, and the temperature of the cosmic background
radiation.

We have seen how  past history can be reconstructed in quantum mechanics 
through retrospective probabilities conditioned on present data. 
We now describe a number of
non-classical features of quantum retrodiction:

\subsection{The Present State is not Enough to Retrodict the Past}
\label{sec:III.A.}

The conditional probabilities for {\it predictions} can be obtained from a
present state.  Define
\begin{equation}
|\Psi_{\rm present}\rangle = \frac{P_{pd(t_0)}|\Psi\rangle}{\Vert
P_{pd(t_0)}|\Psi\Vert} \, .
\label{threeone}
\end{equation}
The probabilities (\ref{twosix}) for future alternatives
conditioned on present data are easily seen to be expressible as
\begin{equation}
p\left(\alpha_nt_n,\cdots,\alpha_1t_1 |pd\, t_0\right) =
\left\Vert P^n_{\alpha_n}(t_n) \cdots P^1_{\alpha_1}(t_1)|\Psi_{\rm
present}\rangle\right\Vert^2\ .
\label{threetwo}
\end{equation}
The vector (\ref{threeone}) is the usual notion of present state of the
system.
In this Heisenberg picture  it is 
constant from the time of the initial condition  up to
the present and then
``reduced'' by the action of the projection representing present
data.  This present state contains all the information necessary
for predicting the future as (\ref{threetwo}) shows.

By contrast no such summary is possible for retrodiction. The
probabilities (\ref{twoseven}) for alternatives conditioned on present
data cannot be expressed in terms of the projections and $|\Psi_{\rm
present}\rangle$ alone.  They require the initial condition $|\Psi\rangle$
in addition to present data $P_{pd(t_0)}$.
Put differently, correct probabilities for the past cannot generally be
constructed simply by running the Schr\"odinger equation backwards in
time from the present state.  A theory of the initial state is required
as well.  In this respect the quantum mechanical notion of state at a
moment of time is very different from the corresponding classical
notion.  
Almost all retrodictions  therefore depend, at least
to some degree\footnote{The trivial exception is the probability
one retrodiction of a history defined by repeating the {\it operator}
$P_{pd}$ over and over at a series of past times. That is a kind
of retrodiction following from the determinism of the Schr\"odinger
equation alone. However, even if $P_{pd}$ represents a quasiclassical 
alternative at the present time it corresponds to nothing like that
at past times. Rather, operators representing quasiclassical alternative
in the past are related to quasiclassical alternatives in the present
by the Heisenberg equations of motion, (\ref{twotwo}).}, on assumptions
about the initial state $|\Psi\rangle$.  We cannot ignore this dependence. Complete ignorance of the initial state would be
represented by an initial density matrix $\rho=I/Tr(I)$ where $I$ is the
unit operator.  But that is an initial condition of infinite temperature
--- patently inconsistent with observations today.

\subsection{The Past is Not Unique}
\label{sec:III.B.}

We naturally think the past defined by
alternative prior histories of the usual quasiclassical variables we are
adapted to observe --- histories of alternative values of densities of
energy, momentum, chemical species, averaged over small volumes, 
 {\it etc}. However, in quantum mechanics
there are many other mutually incompatible possible pasts.

A simple mathematical example may help to illustrate this  point in a
trivial way. Consider a set of projections $\{Q_\alpha\}$ on a basis of
orthogonal states that include $|\Psi\rangle$. The set of histories
\begin{equation}
P_{pd(t_0)} (t_0) Q_{\alpha_1}(\tau_1) Q_{\alpha_2} (\tau_2) \cdots
Q_{\alpha_n}(\tau_n)
\label{threethree}
\end{equation}
is decoherent for any assignment of times
$\tau_1 > \tau_2 > \cdots > \tau_n$ (all earlier than
$t_0$). The conditional probabilities
\begin{equation}
p\left(\alpha_1 \tau_1, \cdots, \alpha_n \tau_n|pd \, t_0 \right)
\label{threefour}
\end{equation}
retrodict a past which is nothing like the usual quasiclassical one.
Indeed the only retrodiction with non-zero probability is for the
history where the $Q's$ are $|\Psi\rangle\langle\Psi|$ repeated at
every time.

We can illustrate this using Schr\"odinger's cat.  Restricting attention
to the usual thought experiment, we may infer from present data
showing the cat to be alive a conditional probability near unity
that the atom did not decay in the past.   However, the conditional
probability is also unity for the past alternative $|\phi\rangle\langle\phi|$
where $|\phi\rangle$ is the state which is a superposition
of states in which the cat is alive and dead.

Indeed an arbitrary number of possible
pasts can be constructed mathematically as follows: Take the vectors
$P_{pd(t_0)} |\Psi\rangle$ and $\bar P_{pd}(t_0)|\Psi\rangle$ where $\bar
P_{pd(t_0)} = I-P_{pd(t_0)}$ (``not the present data'').  
Find a set of mutually orthogonal vectors such
that half of them
add to $P_{pd(t_0)}|\Psi\rangle$ and the remaining ones to
$P_{pd\, (t_0)}|\Psi\rangle$. There are a great many ways of doing
this.  Pick a time $\tau_1$ prior to $t_0$ and write these vectors as
\begin{equation}
P_{pd(t_0)}\ P^1_{\alpha_1}(\tau_1)\, |\Psi\rangle\, ,  \quad
\bar P_{pd(t_0)}\ P^1_{\beta_1}(\tau_1)\, |\Psi\rangle\ .
\label{threefive }
\end{equation}
\noindent where $\alpha_1$ and $\beta_1$ range over half the set of vectors. 
This
can always be done because the projections $P^1_{\alpha_1}(\tau_1)$ 
can always
be found, and, in the Heisenberg picture, a projection operator 
projects on a range of some quantity at any time.  Repeat the process to add
further projections and thereby construct a decoherent set of histories
of prior alternatives that includes present data.  That is a past in
quantum mechanics.  There are a very large number of ways of
interpolating such projection operators and therefore a large number of
quantum pasts.\footnote{Other references discussing the implications
of the multiplicity of quantum pasts are \cite{dEsp87,DK96}.}

Many of these possible quantum pasts are mutually incompatible. That is,
two decoherent sets of histories needed used in retrodiction are not
generally coarse grainings of a finer grained decoherent set of
histories
containing them both. The two pasts in the Schr\"odinger cat example
above are instances.\footnote{Even two alternatives
represented by orthogonal projections can both be retrodicted with
probability one in different quantum pasts for certain present data and
initial and final conditions \cite{Ken97}. Were two such
alternatives to occur in the same past they would be exclusive --- if
one has probability one, the other necessarily has probability
zero.  No logical inconsistency arises from two orthogonal alternatives having
probability one in {\it different} pasts, because, with the relevant
initial and final conditions and present data, there are no decoherent
sets of histories containing both orthogonal projections from whose
probabilities a contradiction could be drawn \cite{GrH97}.}
One can draw past inferences using one set or the other set but not
both\cite{Omn94,Gri96}.

The description of multiple quantum pasts requires care in the use
of ordinary language which is adapted to a single quasiclassical past.
For example, we are
accustomed to say that an event ``happened'' in the past if
its probability conditioned on present data is near unity.  That is 
a statement which requires the history consisting of present data
and the past alternative to be a member of {\it some} decoherent
set of histories, but it is independent of  {\it which} decoherent
set because the value of the probability of a history is the same, 
eq.(\ref{twofour}), in all decoherent sets of which it is a member.
If we retain this meaning of ''happened'' in quantum theory we must accept
mutually incompatible events  
can happen in different quantum pasts.\footnote{A prescription for
avoiding ambiguities of ordinary language is to turn all statements into statements about
quantum mechanical probabilities which are the output of quantum
mechanics.}

What are we to make of these different quantum pasts? It is important to
stress that, in general, we do not expect these to be constructed from
operators representing anything like the alternatives of the usual
quasiclassical realm. Further, in general, we must expect these to be
pasts with distributed probabilities --- no one past history with a
probability near one on the basis of present data and the initial
condition. The possible pasts can thus be expected to differ greatly in
their utility.  We shall return to this below.

\subsection{A Past is Not Necessarily Permanent}
\label{sec:III.C.}

As we move into the future present data changes. Individually we make
more observations and acquire more information over time.  We
also forget or lose access to much. Collectively we expect
information to increase although much is erased or lost such as
Aristotle's {\sl Comedy}.

Thus in general we expect $P_{pd(t)}$ 
 to become increasingly
fine-grained with $t$. 
The decoherence of any particular quantum past is thus at risk. 
A coarse graining of a 
decoherent set is again  decoherent, 
but a fine graining is not necessarily decoherent. 
If a set of past histories no longer decoheres in the presence of
new data, we lose the ability to make these retrodictions. It
is not that the past alternatives become less certain; there
are no probabilities at all. 
A past is therefore not necessarily
permanent in quantum mechanics.

This loss of the past occurs even in familiar laboratory situations.
Consider a Stern-Gerlach thought experiment in which a beam of atoms,
each initially in a superposition of spin states, is separated into two
beams by passing through an inhomogeneous magnetic field, and later
these beams are recombined.  Suppose the atoms are in narrow 
wave packets so that we can
meaningfully speak of the time a single atom is at a particular
position along the beam.  When the atom is in the region of separated
beams, the set of histories defined by alternative
spin  states decoheres. 
(Or indeed at other times since histories consisting of alternatives at
a single time always decohere.) However, a history which includes, in addition to
these alternatives, alternative values of the spin at a time when
the atom is in the region where the beams have been recombined will not
decohere.  The phases between the histories have been recovered in
that present data by recombining the beams. 

As we shall now describe, for quasiclassical pasts, 
it is genuinely or practically
impossible to find present data with which alternative past histories
of values of the usual quasiclassical variables fail to decohere.  As
long as we stick to such alternatives we may proceed into the future
secure in the knowledge that the
quasiclassical past is permanent.\footnote{The decoherence condition
(\ref{twofive}) can be strengthened to ensure the permanence of the
past.  An example is the notion of {\it strong decoherence} 
\cite{GH95}.}

\section{The Quasiclassical Past}
\label{sec:IV}

Classical physics is an approximation to quantum mechanics that is
appropriate for particular coarse grainings and particular initial
conditions such as the one which characterizes our universe.  The
classical past is unique, permanent, and retrodictable from present data
alone --- features which are not general in quantum theory as 
we have discussed.
We now briefly describe how these features are recovered in
 the classical approximation to quantum theory.

The familiar variables of classical physics include averages over
suitable volumes of densities of approximately conserved quantities
such as energy, momentum, chemical species, {\it etc}.  When the volumes are
sufficiently large, sets of alternative histories of such variables
can decohere, and exhibit approximate patterns of classical, deterministic
correlations over time.  The success of classical physics is describing
phenomena over wide ranges of time place and epoch suggests that the
initial condition and Hamiltonian of our universe exhibit a {\it
quasiclassical realm} --- a decohering set of alternative histories of
quasiclassical variables that is maximally refined consistent with the
requirements of decoherence and approximate classical predictability.
Such a quasiclassical realm would extend over most of space and much of
the history of the universe, and is a feature of the universe not of our
choice. As observers in the universe we utilize coarse grainings of this
usual quasiclassical realm to describe everyday experience and
particular experimental situations.

To restrict to past histories of quasiclassical variables is
to beg the question of the uniqueness of the past. However, there is
still a sense in which we may speak of a unique quasiclassical past ---
the past of the usual quasiclassical realm.  Different observers making
use of usual quasiclassical variables for their particular present data
and retrodictions construct pasts which are mutually compatible if
they are all employing coarse grainings of the usual quasiclassical
realm.

It is an open question whether the universe exhibits essentially
distinct quasiclassical
realms of high predictability characterized by variables that
are different from the usual quasiclassical ones
\cite{GH94}.  If,
however, the usual quasiclassical realm is essentially alone in its high
level of predictability that would be another sense in which the usual
quasiclassical past is unique. 
In the following, to avoid cumbersome 
expressions, when we refer ``the quasiclassical realm'' or
``quasiclassical variables'' we mean the usual quasiclassical realm and
the usual quasiclassical variables unless otherwise noted.  

In a restricted class of cases, the equations of motion summarizing 
the temporal regularities of the
quasiclassical realm enable coarse grainings of the quasiclassical
past to be inferred from present data alone without further knowledge of
the initial condition. The history of the solar system, for instance,
can be inferred from the present positions and velocities of its
constituents. However, this kind of retrodiction from present data is
possible only in certain circumstances. 
The effective deterministic equations of motion of the usual quasiclassical
realm
are generally dissipative, chaotic, and dependent on the past history of the
system.\footnote{ 
They are not dependent on future history; classical causality follows from
quantum mechanical causality \cite{GH93a}.} Dissipation and chaos
generally mean that past evolution is unstable.  Dependence on past
history
means that present data is not enough to retrodict. Further,
thermal and quantum noise arising from coarse
graining can cause deviations from classical 
predictability, both for prediction and retrodiction.  Only when the effects
of dissipation, chaos, history dependence, and noise are sufficiently
small can we use equations of motion to retrodict the usual
quasiclassical past from present data alone.
Unlike classical physics where coarse grainings and effective
equations arise only out of ignorance, coarse graining is 
inevitable in quantum mechanics. Coarse graining is necessary for
decoherence, and for the quasiclassical realm, coarse graining
beyond that is necessary to achieve the approximate predictability 
in the face of the noise that typical mechanisms of decoherence
produce \cite{GH93a}.

It is impossible or hopelessly impractical to
recover the phase information in present data that would lead to a
failure of decoherence of the usual quasiclassical past.  
We mentioned that, for an intergalactic dust grain
initially in a superposition of positions, alternative histories of
subsequent position decohered because of their interaction with cosmic
background photons. If later data includes the states of all scattered
photons, then the past histories of position of the dust grain
cannot be retrodicted from it
--- the relevant set of histories of position
 will not decohere.  Of course, it is hopelessly impractical to
recover such data. Some $10^{11}$ photons scatter off the dust grain
every second.  Indeed, if we did not prepare ahead of time to recover
the information, its too late in principle.  It left earlier at the speed of light.
If any of those photons went down a black hole they are similarly
irretrievably lost. 

In science fiction we might imagine that
even now intelligent aliens could be poised to recover the phases
between past histories that we retrodict from present data.  From this
data they will not be able to retrodict the same past as we do. We,
however, are unlikely to survive the process of collecting this data to
compare notes.

In the quasiclassical past one can say
with the poet, 
``The Moving Finger writes; and having writ, Moves on:
nor all thy Piety nor Wit Shall lure it back to cancel half a Line, Nor
all Thy tears wash out a Word of it.''  That is not necessarily
the case for all quantum mechanical pasts. 

\section{Why Are there More Records of the Past than the Future?}
\label{sec:V}

In the quasiclassical realm, 
present data contain many more records of the past than of the
future.  In this paper a ``record'' is not presupposed to be of
the past. A record is an an alternative in the present that is
correlated with high probability with another alternative {\it in the past or
future}. That is, present data contains a record of an
alternative $\alpha_0$ if  $p(\alpha_0,
t|pd, t_0)\approx 1 $  irrespective of the relation between $t$ and $t_0$.
A recorded alternative is thus either a near certain prediction
or retrodiction from present data. For example,
the configuration of ink read 55BC in many texts (and 55BCE in others) 
is correlated
with the Roman invasion of Britain. Certain configurations of neurons
are correlated with our past experiences.  We can also have records of
the future. A table of the eclipses of the moon in AD2010 
is just as
much a record of that time as a table of eclipses in AD1000 is of its. The present
position and velocity of the earth in its orbit around the sun are 
a record of its future positions as well as of its past ones. 
However, the preponderance of records seem to be of the past.
That time asymmetry, like all others, can only be explained by
the particular features of the initial condition $|\Psi\rangle$.
In this section, we shall attempt to identify these features. 
We confine our attention entirely to the usual quasiclassical
realm. 

The second law of thermodynamics is the reason that present data
contains more records of the past than the future. The Jaynes
construction may be used to associate an entropy with
with every decoherent set of alternatives in quantum
mechanics.\footnote{ If that is not familiar see
{\it e.g.} \cite{Ros83}. By extending the original Jaynes discussion 
an entropy can be defined for sets of histories \cite{GH90a}. } 
For example, the entropy of the coarse-grained set of histories
consisting of alternatives $\{P_\alpha (t)\}$ at a single moment
of time is:
\begin{equation}
S\left(\{\alpha\}, t \right) =  - \sum\nolimits_\alpha
p\left(\alpha, t \right) {\rm log}\ p\left(\alpha, t
\right)
+ \sum\nolimits_\alpha p\left(\alpha, t \right) {\rm log} 
\ Tr\left[P_\alpha(t)\right]\ .
\label{fourX}
\end{equation}
where $p(\alpha, t)=||P_\alpha(t)|\psi\rangle||^2$ are 
the probabilities determined from the
initial condition.  Now follow 
this entropy with time keeping the coarse-graining fixed. When the $P's$
are projections on ranges of {\it quasiclassical variables} this 
entropy is low initially for the initial condition of our universe
and therefore has a general tendency to increase afterwards.\footnote{
While low, the entropy is not as low as it could be. At the time
of decoupling, matter and radiation were in near thermal equilibrium
with an entropy of about $10^{80}k$ inside the region visible from today.
By contrast, the geometry of the early universe is as ordered as it 
could be. As Penrose has stressed \cite{Pen79}, ``gravitational
entropy'' is small compared to the maximal value of $10^{120}k$
it could have if all that matter were in a black hole. The second law
is a consequence of this initial geometrical order and the
attractive nature of gravity.

For the entropy of the initial state to be near zero with respect to
quasiclassical coarse grainings, it would have to predict a single
configuration of quasiclassical variables with probability near unity
[{\it cf.} (5.1)]. The subsequent approximate deterministic evolution of that
configuration means that it must necessarily encode all the complexity
of the quasiclassical realm we see about us. Such a near zero entropy
initial condition would therefore be
incomprehensibly complex. Rather the initial condition of our universe
seems to be as {\it simple} as possible in a crude, intuitive sense: Matter
in near thermal equilibrium which is maximum entropy for it, and 
highly ordered geometry which is minimal entropy for it because of the
attractive nature of gravity. Quantum mechanical initial conditions which
incorporate this simplicity ({\it e.g.}\cite{HH83}) predict distributed
probabilities for histories of the quasiclassical realm which is
why there are few initial records of the future. }
 That is
the second law of thermodynamics and the origin of the thermodynamic
arrow of time.  As Boltzmann put it \cite{Bolquote}, 
 ``The second law of thermodynamics can be proved from
the [time-reversible] mechanical theory if one assumes that the present
state of the universe\dots started to evolve from an improbable
state.''

Similar second law behavior is expected for the entropy  
of a set of alternatives $\{P_\alpha (t)\}$ conditioned on present
data. This is given analogously to (\ref{fourX}) by 
\begin{eqnarray}
S\left(\{\alpha\}, t | pd, t_0\right)& = & - \sum\nolimits_\alpha
p\left(\alpha, t | pd, t_0\right) {\rm log}\ p\left(\alpha, t|pd,
t_0\right)\nonumber\\
&+& \sum\nolimits_\alpha p\left(\alpha, t | pd\, t_0\right) {\rm log} 
\ Tr\left[P_\alpha(t)\right]\ .
\label{fourone}
\end{eqnarray}
where the conditional probabilities $p(\alpha,t|pd,t_0)$ are constructed
as described in Section IIC.
For a generic quasiclassical coarse graining this conditional
entropy should increase towards the future and decrease towards the
past. 

If present data contains a record of a particular alternative $\alpha_0$,
then the conditional probability for it is unity and the conditional
probability is zero for all other alternatives in any decoherent set
containing it.  
The entropy of such a set is then
\begin{equation}
S\left(\{\alpha\}, t|pd, t_0\right) = {\rm log} \, Tr \left[P_{\alpha_0}
(t)\right] \, . 
\label{fourtwo}
\end{equation}
This is {\it independent of time} as a consequence of the unitary
evolution of the $P$'s (\ref{twotwo}) and the cyclic property of the
trace. That is consistent with the second law. The second law thus
does not rule out particular near certain predictions or retrodictions.
For example, the entropy of a set of alternatives describing the positions
of the earth and the moon to the accuracies current in celestial
mechanics increases hardly at all in the near term. That is
because their motion is essentially deterministic. 
 A table of future eclipses is an example of a
future record of such alternatives.\footnote{ Note that the effort to
compute the table does not  enter the argument, nor should it. 
The present records of positions of the  earth-moon
system are correlated with their  future positions whether or
not these records are the input to explicit computation. 
{From} the point of view of entropy, computation is tantamount to copying and
does not necessarily increase the entropy \cite{BenXX} unless 
outputs of a reversible computation are erased in accord with the 
Landauer's principle \cite{Lanprinc}. Computation is typically
required in the process of prediction to generate {\it useful} records
of future alternatives. In the computation of a table of eclipses
an account of complex astronomical observations is converted to 
a more useful table of dates
in the amended Gregorian calendar, both equally being records of the
future astronomical events.}

The question, however,  is not whether individual records are consistent with
the second law, but whether present data contains more near certain
retrodictions than near certain predictions. The second law implies
that it does. Pick a set of quasiclassical alternatives
$\{P_\alpha(t)\}$.   Their  entropy is generically greater at times
after the time of present data than it is for times before.
Therefore, generically the probabilities $p(\alpha,t)$ are 
more distributed (less concentrated on one certain alternative) when $t$ is to
the future of the time of present data than when it is to the past of
it. That means a given quasiclassical coarse graining is more likely
to contain a near certain retrodiction of the past than a near certain
prediction of the future.

This is in accord with the intuitive understanding that many
records originate in irreversible processes. An impact crater on the
moon, an ancient fission track in mica, or the arrangement of ink
on this page are all examples of records in which the entropy rose
significantly when they were created. Records created by irreversible
processes must necessarily be records of the past to be consistent
with the second law.

\section{The Utility of History}
\label{sec:VI}

We reconstruct the past to help predict the future.  Certain
regularities in the universe that can be inferred from present data can
be understood as arising from historical events.  This understanding is
useful for predicting future regularities and that is the utility of
history.  In this section we shall amplify on this theme.  We shall
briefly discuss the nature of regularities, the process of historical
explanation, quantitative measures of
explanation, the role of theoretical input to that process, why
explanations are mostly in the past, and the implications of many 
quantum pasts for 
historical explanation in quantum mechanics.

\subsection{The Origin of Regularities}
\label{subsec:VI.A}

Science is concerned with identifying regularities in the universe and
exploiting these in the process of prediction. 
The laws which govern the regularities exhibited by all
physical systems --- without exception, without qualification, and
without approximation --- are the fundamental laws of physics. 
The present theoretical viewpoint is
that there are two fundamental laws --- a unified theory of the dynamical
interactions, perhaps heterotic superstring theory
or a generalization of it, and a theory of the
initial quantum state of the universe.

Beyond the universal laws, science, including physics, aims a
identifying and exploiting the regularities of specific physical
systems. To give just a few examples: physics is concerned with the
regularities of atoms, stars, and fluid flows; chemistry with
the regularities of specific molecules; geology with the regularities of
a specific planet;  and the social sciences with the regularities
exhibited by human beings --- both collectively and individually.
Specific classes of systems exhibit many more regularities than those
implied by the fundamental laws of physics.  What is the origin of the
regularities peculiar to specific subsystems of the universe? Evidently
they do not arise from the fundamental laws, as these govern
only the regularities exhibited by all systems.

Every prediction in science can be regarded as  a conditional
probability in quantum cosmology. The regularities of specific systems
can therefore only arise from the conditions entering the
probabilities for these
regularities.  In some cases merely specifying the system is enough. To
predict the emission spectrum of an isolated atom it is sufficient to
specify that the atom is isolated in a certain excited state.  For many
other systems, however, the regularities arise from processes and events
that have occurred over the course of the system's history\footnote{See,
{\it e.g.} \cite{Gel94} for a more detailed discussion.} --- some
deterministic, some accidental. That is certainly the case for the
regularities exhibited by the geology of the earth or by particular
biological species that are the products of billions of years of chance
mutation and natural selection. Even
physics, however, is concerned with regularities which arise from
historical events. The large scale
distribution of galaxies in the visible universe is most
succinctly understood as the outcome of some 15 billion years of evolution
of primordial fluctuations by the action of gravitational attraction.
This historical understanding from present data can be used to predict
regularities in more detailed data yet to be obtained.  For example, by
reconstructing the past history of the elements,  we can
predict the abundances in  meteorites and stars yet to be observed. In
the following we describe this process of historical explanation in
quantum mechanics.

\subsection{Historical Explanation}
\label{subsec:VI.B.}

{From} present data that includes most  texts giving 55BC as the date of
the Roman invasion of Britain, we may infer that Caesar did invade
Britain in 55BC. That event, together with inferences from present data
on the validity of the texts, is a {\it historical explanation} of the
regularity in present data that most discussions of the Roman invasion give
55BC as its date.  From this explanation one can predict that texts
yet to be discovered will also give the date 55BC when describing the
Roman invasion of Britain. (Of course, both inference and prediction are
probabilistic; there is a probability that individual texts are
forgeries, or contain mistakes,
or that the ink on their pages made a quantum mechanical transition
from a configuration spelling a different date.) The present coherence
among texts on the date of the Roman invasion is an example of a ``frozen accident'' in the evocative
terminology of Gell-Mann \cite{Gel94} --- ``chance events of which the
particular outcomes have a multiplicity of long term consequences all
related by their common ancestry''. 

In principle the same prediction could be made from the present data
itself --- from the physical description of the texts and the
configuration of ink molecules on their pages. However, it is evidently much
easier to start from the event in the past.  The reason is that present
data contains much information that is irrelevant for this particular
future prediction. It contains the specifications of the texts --- their
location, content, typeface language, number of pages, authors, {\it
etc.~etc.}, --- many details to which the prediction of 55BC in future texts is
not very sensitive. Of course, all that data is relevant for other
predictions, those referring to the subsequent evolution of the texts
for instance.  But the reconstruction of past history permits the focus
on the {\it essential} features of the regularity
 that can be inferred from present data
for the prediction of the character of future texts.

The above discussion can be given both more generally and more concretely. Let
$P_{pd(t_0)}$ be the projection on present data held at time $t_0$, and
let $\{P_\alpha(t_0)\}, \alpha=1,2,\cdots$ be a set of alternatives in
which $P_{\alpha_0}(t_0)$ is a coarser projection than $P_{pd(t_0)}(t_0)$
describing a feature, say a regularity, in present data.  Histories
\begin{equation}
C_{\beta_0} = P^1_{\beta_{1,0}}(\tau_1) \cdots P^n_{\beta_{n,0}}(\tau_n)
\label{sixone}
\end{equation}
are inferences from present data if
\begin{equation}
p(\beta_0|pd) \approx 1\ .
\label{sixtwo}
\end{equation}
(In this section we will suppress the time labels of alternatives for
compactness.)
If among the possible inferences  from present data,
 we can find one $\beta_0$ for which
\begin{equation}
p(\alpha_0|\beta_0) \approx 1 \, ,
\label{sixthree}
\end{equation}
then we say that $\beta_0$ is an explanation of $\alpha_0$.

In the history of the Roman invasion of Britain, $pd$ contain all the
details of present texts, $\alpha_0$ is the coarse grained alternative
that most texts report 55BC for the invasion, and $\beta_0$ is the chain 
of events in 55BC describing the invasion. In explaining the
primordial abundance of the elements, $pd$ includes the details of
telescopic observations of the spectra from old, metal-poor stars,
$\alpha_0$ is the alternative that these all show approximately 75\%
H and 25\% He, and $\beta_0$ is the history of big bang
nucleosynthesis in the early universe.

To consider probabilities such as (\ref{sixtwo}) and (\ref{sixthree})
at all, the alternatives $\beta_0$ and $\alpha_0$ must be members
of decohering sets of histories. 
As throughout this section, we assume  the decoherence of the 
relevant sets of histories. We generally do not need to mention
the set because the probabilities of specific alternatives such as 
$\beta_0$ and $\alpha_0$ are the same in all decoherent sets, being 
given by (\ref{twofour}).  

There will generally be many possible explanations of a given
alternative $\alpha_0$.  To give a trivial example, merely
repeating present data or evolving it by the Schr\"odinger equation
(\ref{twotwo}) to a different time would give a $\beta_0$ satisfying
(\ref{sixtwo}) and (\ref{sixthree}).  
However, it is not
explanation in general which is of interest, but rather explanation 
which is simpler than present data.

Quantitative measures of the simplicity of explanations depend
on the context that is assumed in comparing them. The simplest
measure is the information content (AIC) of the histories  $\beta_0$
satisfying (\ref{sixtwo}) and (\ref{sixthree}). 
The smaller the AIC, the simpler the
explanation. The notion of AIC was defined
some thirty years ago by Kolmogorov, Chaitin, and Solomonoff \cite{LV93}
(all
working independently) and has been employed in the
definition of augmented thermodynamic entropy \cite{Zur89}, the
effective complexity of physical systems \cite{GL96}, and 
measures of classicality \cite{GH95}. For a string of bits $s$ and a
particular universal computer $U$, the AIC of $s$, written $K_U(s)$, is
the length of the shortest program that will cause $U$ to print the
string and then halt.  A history like $\beta_0$ , eq.(\ref{sixone}),
may be described by giving
a description of its string of projection operators and their times.   
A description of a projection operator entails describing the 
operator whose eigenvalues are projected upon in terms of fundamental
fields and specifying the range of these eigenvalues. 
If $s$ is a description of such a history 
then the AIC of the string can be
regarded as the AIC of the history. There will generally
be many physically equivalent descriptions of a history in terms of fields
\cite{GH94}.
For example, physically equivalent explanations are obtained
by choosing different times for the operators in (\ref{sixone})
and describing the operators by fields at that time. However, the AIC's
of these equivalent histories may differ greatly.  
 We therefore take the AIC of the history,
$K_U(\gamma)$ to be the minimum over all such descriptions.\footnote{
More accurately, the AIC is defined on the physical equivalence class
of descriptions of $\alpha_0$ and $\beta_0$ \cite{GH94}.}  
The AIC, $K_U(\beta_0 )$, of an explanation $\beta_0$ 
then provides a quantitative way of distinguishing between different
explanations by their relative simplicities. The $\beta_0$ with the
smallest AIC is the simplest explanation. Since $\alpha_0$ trivially
satisfies (\ref{sixtwo}) and (\ref{sixthree}) it is an explanation
of itself, and the simplest explanation of $\alpha_0$ will have
an AIC lower than that of $\alpha_0$ itself. 

Not every explanation, simple or not,  is useful for prediction. The probabilities
$p(\gamma|pd)$ of an arbitrary set of alternatives $\{P^1_\gamma(t_1)\}$
need be neither more nor less distributed than the probabilities
$p(\gamma|\beta_0)$ conditioned on an explanation of part of that data.
However, if the explanation $\beta_0$ captures the
essential features of present data that are relevant for the
alternatives $\{P^1_\gamma(t_1)\}$, then their prediction may be
simplified.  That will be the case when the probabilities of the
$\{P^1_\gamma(t_1)\}$ are independent of the part of present data not
explained by $\beta_0$ in the sense that
\begin{equation}
p(\gamma|pd, \beta_0) = p(\gamma|\beta_0)\ .
\label{sixfive}
\end{equation}
For such alternatives, the predictions from $\beta_0$ agree with those
from present data
\begin{equation}
p(\gamma|pd)=p(\gamma|\beta_0)\ ,
\label{sixsix}
\end{equation}
because of
\begin{equation}
p(\gamma|pd) = \Sigma_\beta p(\gamma|pd, \beta)p(\beta|pd)
\label{sixseven}
\end{equation}
and eqs (\ref{sixtwo}) and (\ref{sixfive}). 
When the explanation $\beta_0$ is a simpler input to the
prediction of the $\{P^1_\gamma(t_1)\}$ than all of present data,  the
prediction of the future is simplified by historical
explanation.

The above discussion does not assume that explanations must always be in
the past, and indeed they need not be.  The present position and
velocity of the earth in its orbit around the sun might just as well be
explained by some future position and velocity towards which it is
heading as by any past position and velocity from which it came.  However,
the second law argument of the Section III implies that many
explanations will be found in the past.  Eq (\ref{sixtwo}) shows that
present data contains a record of the explanation $\beta_0$, and most,
but not all, records are of the past because of the second law.  
Eq (\ref{sixthree}) shows that $\beta_0$ is a record of the alternative
$\alpha_0$ to its {\it future}. However, $\alpha_0$ is typically only a
small part of present data --- a coarse graining of it.  It is more consistent
with the general tendency of entropy to increase to have the records
arranged in this way rather than the opposite time order. That is why
the past is the most promising place to look for explanations.

There are no past explanations without theoretical input beyond present
data. The explanation of the primordial abundance of the elements by big
bang nucleosynthesis entails a number of assumptions about the nature of
the early universe --- its approximate homogeneity and isotropy, for
instance.  Those theoretical assumptions may be most properly understood
as hypotheses about the universe's initial condition.  Indeed we know
from the discussion in Section IIIA that a theory of the initial
condition is required for any retrodiction from present data in quantum
mechanics.

As we have mentioned earlier, there will generally be many explanations
--- many $\beta_0$ satisfying eqs (\ref{sixtwo}) and (\ref{sixthree})
--- of a given part of present data.  These will differ in their
simplicity and their utility for future prediction.  There is no
requirement that these different explanations be mutually compatible.
They could belong to different quantum pasts in the sense of Section
IIIB. 
With its many quantum pasts, the quantum theory of closed systems
has more possibilities for historical explanation than it would if the
theory were somehow restricted to a single decoherent set of histories.
Mere decoherence in the presence of present data and the initial
condition are all that is required to have a set of histories defining
a quantum past. But to be useful for historical explanation, the 
probabilities for records and explanations in (\ref{sixtwo}) and
(\ref{sixthree}) must be near unity. The approximate determinism
of the usual quasiclassical realm makes it more likely that 
these conditions will be satisfied in the usual quasiclassical 
past for usual quasiclassical present data than in other possible
pasts. That is why the usual
quasiclassical past is the most promising place to look for
historical explanation. 
But we should
not ignore the possibility of explanation through the many other pasts
presented to us by quantum mechanics.

\section{WHAT ABOUT US?}
\label{sec:VII}

\subsection{Present, Past, and Future}
\label{subsec:VII.A}

``Observers'' or ``measurements'' have no fundamental role in the
quantum mechanics of closed systems.   Of course, ``observers'' 
can be described as special physical systems (typically complex ones) within
the universe, and ``measurements'' can be described as special
interactions between subsystems of the universe. ``Observers'' and
``measurements'' are of special interest to us because that's what we 
are and that's how we learn about the universe. 
How then do information gathering and utilizing systems (IGUSes)
such as ourselves distinguish the present moment,  its future, and its past? 
The author is not equipped to speculate on the detailed operation of the human
brain.  However, it seems likely that mechanical systems such as
computers could be constructed and programmed to draw similar 
distinctions between past, present, and future. We proceed 
by discussing how such a system could operate.  
We restrict attention to IGUSes whose description and observations
are coarse-grainings of the usual quasiclassical realm. 
The discussion in this section is necessarily more suppositional than
that in the rest of the paper, and the reader should take note of that.
Nevertheless, it concerns questions that can be seriously addressed
in science, even if now only by conjecture and not calculation, as here.

An IGUS holds a set of records.  As time changes (in either
direction!) this set of records also changes. New records appear and
others disappear or are altered.  The records can
therefore be approximately time ordered. 
How accurate the
records are at any time, how long and faithfully they persist, and how
accurately they can be time ordered depends on the particular IGUS.
Everyday experience shows considerable variation in these qualities from
IGUS to IGUS.

If the records of an IGUS are examined at any one time 
we expect some to be correlated with external events near that time.
That is the present input, and these records constitute the IGUS's 
notion of the present. To the extent that the IGUS's records are
acquired in irreversible processes, we expect most of the rest 
to be records of the {\it past} as a consequence of the second law
as discussed in Section III. That is why we say we remember the
past and not the future.\footnote{The author is not alone in connecting the ``psychological
arrow of time'' to the thermodynamic arrow of time, see
{\it e.g.}~\cite{Pen79,Haw87}.} However, an IGUS is not prohibited 
from possessing a record
of the future.  It might have memorized a table of future eclipses.  It
might have correctly conjectured the motion of a fly or the movement 
of tomorrow's stock market. 
An IGUS must predict to function, which means creating useful
approximate records of the future. As the above examples illustrate, such records of the future are
typically distinguishable  from records of the past as 
the outcomes of {\it computation} involving other records as input,
mostly records of the past. That is how we can say that
we predict the future but remember the past.

Thus, like all other time asymmetries, the subjective distinction between
past, present, and future is traceable to properties of the universe's 
initial condition  --- in this case its low entropy in
terms of usual quasiclassical variables. All other aspects of
the phenomena loosely called the ``psychological arrow of time''
presumably
have similar explanations. For example, we have the impression that 
we can control the future, but the past is over and done with. What
is meant by ``control'' is the amplification of a pattern of 
weak electrical activity in the brain to large irreversible effects
correlated with that pattern. The second law means that such effects
are much more likely to be in the future of the pattern of activity
than to its past. 

\subsection{Which Past Do We Remember?}
\label{subsec:VII.B}

Which of the
many possible quantum pasts do we remember?  As IGUSes we are adapted to
observe and record alternative values of quasiclassical operators.  It
is no surprise that we individually remember quasiclassical pasts if our
records are almost entirely of quasiclassical alternatives.  Further,
these individual quasiclassical pasts are all coarse grainings of the
usual quasiclassical realm which is why they are in  approximate
agreement.  
However, at a deeper level, we may ask: ``What is the
reason for our focus on quasiclassical alternatives?'' Our particular
properties as physical systems in the universe can only be understood in
the context of our evolution within it.  The usual quasiclassical realm  
of decoherent histories has a high level of predictability in time.
Such properties make it plausible
that IGUSes described in terms of usual quasiclassical variables evolved
to make use of coarse grainings of the usual quasiclassical realm 
because its relative predictability makes it adaptive to do so in
comparison with more quantum mechanical alternatives they might have
used.  Probabilities relevant for that adaptation are in principle
computable from quantum mechanics although well beyond our ability to do
so at present.\footnote{For further discussion see \cite{GH94}.}

\subsection{The Marvelous Moment ``Now''}
\label{subsec:VII.C}

To function, an IGUS employs a self-referential, evolving model or
schemata of itself and its external environment \cite{Gel94}. Such a
model necessarily distinguishes records of present events from ones in
the future and past as described above  
  and, in the case of human beings, devotes a
significant amount of conscious focus to records of the most recent
input. The author knows of no physical reason why a device could not be
constructed which gives equal conscious focus to records of an hour ago
as well as of the present.   Plausibly it has been adaptive to
concentrate on the present and near past.  However, our
preponderance of attention to the present does not mean that ``now'' is somehow
distinguished in physics.  Neither is it
evidence that for the more radical but problematical suggestion that
physics can be formulated entirely at one marvelous moment 
``now''\cite{now}. To
give just one reason, the moment ``now'' is only approximately defined
to the accuracy of some processing time in the brain and certainly not
to the accuracy of Planck time, $10^{-43}$sec, on which spacetime can be
defined.\footnote{For more reasons see, {\it e.g.} \cite{Har91a}, Section
V.1.2.}  ``Now'' is a moment in a 
schemata of an IGUS, not a preferred
surface in the four-dimensional spacetime continuum.

\section{Conclusion}
\label{sec:VIII}

The quantum mechanical past is a country of very different appearance
from the past of classical physics. Present data, no matter how refined, are
not enough for any retrodictions without being augmented by a theory of
the initial condition. Even then, there are many mutually incompatible
pasts, each a possible source of historical explanation. As new data is
acquired none of these possible quantum pasts need continue to be
retrodictions --- a quantum past is not necessarily permanent.  The familiar
properties of uniqueness, permanence, and retrodictibility from present
data alone, that are characteristics of the classical past emerge in quantum theory, not
generally, but as approximations appropriate to particular
coarse-grainings and particular initial conditions such as that for our
universe.

The multiplicity of mutually incompatible sets of alternative decoherent
histories that can be employed for prediction and retrodiction in the
quantum mechanics of closed systems has motivated some to search for
additional principles that would restrict the available sets in some
way \cite{Ken97a,Ana97} and thereby necessarily recover properties of the quasiclassical past that
are not available generally in quantum theory. Sum-over-histories 
formulations \cite{SOHQM}, strong decoherence \cite{GH94},
and ordered consistency \cite{Ken96}
can all be seen as efforts in this direction. With 
such restrictions the properties of quantum past could be different and
perhaps closer to those of classical physics.  It is important to
stress, however, that,  even without such  additional principles, the
quantum theory of closed systems is logically consistent, consistent
with experiment as far as is known, and applicable for prediction and
retrodiction to the most general physical systems. 

\acknowledgments

Thanks are due to T.~Brun, R.~Griffiths, J.~Halliwell, D.~Page, and W.~Unruh
for critical readings of the manuscript. 
The author has benefited from discussions with Murray Gell-Mann on many
of the subjects considered in this paper over a
long period of time.
This work was supported in part by NSF grants PHY95-07065 and
PHY94-07194.


\begin{references}

\bibitem{Omn94} For a survey see, R.~Omn\`es, {\it Interpretation of Quantum Mechanics},
(Princeton University Press, Princeton, 1994).


\bibitem{TNQMsum} Y.~Aharonov,  P.~Bergmann,  and J.~Lebovitz,
{\sl Phys.~Rev.~B}, {\bf 134}, 1410 (1964);
R.B.~Griffiths, {\sl  J.~Stat.~Phys.}, {\bf 36}, 219 (1984);
M.~Gell-Mann and J.B.~Hartle, {\it Time Symmetry and
Asymmetry in Quantum Mechanics and Quantum
Cosmology},
in {\sl Proceedings of the NATO Workshop on the Physical
Origins of Time Asymmetry, Mazag\'on, Spain, September 30-October4,
1991}
ed. by J. Halliwell, J. P\'erez-Mercader, and W. Zurek, Cambridge
University Press, Cambridge (1993); gr-qc/9309012.

\bibitem{Laf93} R.~Laflamme,  {\sl  Class.~Quant.~Grav.}, {\bf 10}, L79,
1993, gr-qc/9301005.

\bibitem{Cra95} D.~Craig,
{\sl Annals of Physics (NY)}, {\bf 251}, 384 (1995), gr-qc/9704031. 

\bibitem{JZ85} E.~Joos  and H.D.~Zeh, {\sl Zeit.~Phys.}, {\bf B59}, 223
(1985).

\bibitem{Har91a} J.B.~Hartle, {\it The Quantum Mechanics of Cosmology},
in {\sl
Quantum Cosmology and Baby Universes:  Proceedings of the 1989 Jerusalem
Winter
School for Theoretical Physics}, ed.~by ~S.~Coleman, J.B.~Hartle,
T.~Piran,
and S.~Weinberg, World Scientific, Singapore (1991), pp.~65-157.

\bibitem{dEsp87} B. d'Espagnat, {\sl Phys. Lett. A}, {\bf 124}, 
204, (1997) and the reply by R.~Griffiths, {\sl Found. Phys.}
{\bf 23}, 1601 (1993).

\bibitem{DK96}  H.F.~Dowker and A.~Kent, 
{\sl J. Stat. Phys.} 82, 1574, (1996),
gr-qc/9412067..

\bibitem{Ken97} A.~Kent, {\sl Phys.~Rev.~Lett.}, {\bf 78}, 2874 (1997),
gr-qc/9604012; G.~Peruzzi and A.~Rimini, quant-ph/9710003.

\bibitem{GrH97} R.B.~Griffiths and J.B.~Hartle, {\it Comment on
``Consistent Sets Yield Contrary Inferences in Quantum Theory''};
gr-qc/9710025.

\bibitem{Gri96} R.B.~Griffiths, {\sl Phys. Rev. A} {\bf 54}, 2759
(1996), quant-ph/9606004.


\bibitem{GH95}M.~Gell-Mann and J.B.~Hartle, {\it Strong Decoherence},
to be published in the {\sl Proceedings of the 4th Drexel Symposium on
Quantum Non-Integrability --- The Quantum-Classical Correspondence,
Drexel University, September 8-11, 1994}, ed. by. D.-H. Feng and B.-L.
Hu, International Press, Boston/Hong-Kong; gr-qc/9509054.


\bibitem{GH94} M.~Gell-Mann and J.B.~Hartle, {\it Equivalent Sets of
Histories and Multiple Quasiclassical Domains}, 
gr-qc/9404013.

\bibitem{GH93a} M.~Gell-Mann and J.B.~Hartle,  {\sl Phys.~Rev.~D},
{\bf 47}, 3345 (1993); gr-qc/9210010.
An abbreviated account of this paper is given in:
J.B.~Hartle, {\it Quasiclassical Domains in a Quantum Universe}
 in {\sl Proceedings of the Cornelius  Lanczos International
 Centenary Conference}, North Carolina State University,
December 1992, ed.~by J.D.~Brown, M.T.~Chu, D.C.~Ellison, R.J.~Plemmons,
SIAM, Philadelphia, (1994), gr-qc/9404017.

\bibitem{Ros83}  R.D.~Rosenkrantz, ed.{\it  ~E.T.~Jaynes:
Papers on Probability Statistics and Statistical Mechanics}, D.~Reidel,
Dordrecht (1983).

\bibitem{GH90a} M.~Gell-Mann and J.B.~Hartle, {\it Quantum Mechanics in
the Ligh
t of Quantum Cosmology}, in {\sl Complexity, Entropy, and the Physics of
Informa
tion, SFI Studies in the Sciences of
Complexity}, Vol.  VIII, ed. by W. Zurek,  Addison Wesley, Reading, MA
or in {\sl Proceedings of the 3rd
International Symposium on the Foundations of Quantum Mechanics in the
Light of
New Technology} ed.~by S.~Kobayashi, H.~Ezawa, Y.~Murayama,  and
S.~Nomura,
Physical Society of Japan, Tokyo (1990).

\bibitem{Pen79} R.~Penrose, {\it Singularities and Time Asymmetry}
in  {\sl General Relativity: An Einstein
Centenary Survey},  ed.~by S.W.~Hawking and W.~Israel,  Cambridge
University
Press, Cambridge (1979).

\bibitem{HH83} J.B.~Hartle and S.W.~Hawking, {\sl Phys. Rev. D}, {\bf
28},
2960 (1983).

\bibitem{Bolquote} L.~Boltzmann, 
{\sl Ann.~Physik},
{\bf 60}, 392, (1897), as translated in S.G.~Brush, {\it Kinetic
Theory}, (Pergamon Press, New York, 1965). 

\bibitem{BenXX} C.H.~Bennett,
{\sl IBM Jour. of Res. and Development} {\bf 17}, 525 (1973),
and, {\sl Int.~J.~Theor.~Phys.}, {\bf 21},
905-940 (1982).


\bibitem{Lanprinc} R. Landauer, 
{\sl IBM Journal of Research and Development}, {\bf 5}, 
183 (1961); and {\sl Nature}, {\bf 355}, 779 (1988).  


\bibitem{Gel94} M.~Gell-Mann, {\it The Quark and the Jaguar},
(W.~Freeman, San Francisco, 1994).

\bibitem{LV93} M.~Li and P.~Vit\'anyi, {\it Introduction to Kolmogorov
Complexity and Its Applications}, Springer, New York (1993).

\bibitem{Zur89} W.~Zurek, {\sl Phys.~Rev.~A}, {\bf 40}, 4731 (1989).

\bibitem{GL96} M.~Gell-Mann and S.~Lloyd, {\sl Complexity}, {\bf 2},
44 (1996).

\bibitem{Haw87} S.W. Hawking, {\sl New Scientist}, {\bf 115}, 46
(1987).

\bibitem{now} For some recent versions of this see {\it e.g.} 
S.~Coleman, {\it Quantum Mechanics in Your Face} 
(unpublished lecture) and D.~Page, {\it Sensible Quantum Mechanics: Are
Probabilities Only in the Mind?}, quant-ph/9506010. 

\bibitem{Ken97a} See for example the review by A.~Kent in this volume.

\bibitem{Ana97} C.~Anastopoulos, {\it On the Selection of Preferred
Consistent Sets}, quant-ph/9709051. 

\bibitem{SOHQM} See, {\it e.g.} 
J.B.~Hartle, {\it Spacetime Quantum Mechanics and the
Quantum Mechanics of Spacetime} in {\sl
Gravitation and Quantizations}, Proceedings of the 1992 Les Houches
Summer School, edited by B. Julia and J. Zinn-Justin, Les Houches Summer
School Proceedings Vol. LVII (North Holland, Amsterdam, 1995),
 gr-qc/9304006; R.~Sorkin,  
  {\it Quantum Measure Theory and its Interpretation}, in
      D.H.~Feng and B.-L.~Hu (eds.),
      {\sl Proceedings of the Fourth Drexel Symposium on Quantum
         Nonintegrability: Quantum Classical Correspondence},
      (International Press, 1996), gr-qc/9507057. 

\bibitem{Ken96} A.~Kent,  {\it Quantum Histories and Their
Implications}, gr-qc/9607073.

\end{references}
\end{document}